\documentclass[aps,prl,onecolumn,superscriptaddress,showpacs]{revtex4}

\usepackage{amsmath}    
\usepackage{amssymb}    
\usepackage{graphicx}   
\usepackage{epsf}
\usepackage{epsfig}
\usepackage{subfigure}
\begin{document}

\title{New Experiments for Spontaneous
Vortex Formation in Josephson Tunnel Junctions}
\thanks{submitted to Phys. Rev. B.}

\author{R.\ Monaco}
\affiliation{Istituto di Cibernetica del C.N.R., 80078, Pozzuoli,
Italy and Unita' INFM-Dipartimento di Fisica, Universita' di
Salerno, 84081 Baronissi, Italy.}\email
{roberto@sa.infn.it}
\author{M.\ Aaroe and J.\ Mygind}
\affiliation{Department of Physics, B309, Technical University of
Denmark, DK-2800 Lyngby, Denmark.} \email{myg@fysik.dtu.dk}
\author{R.\ J.\ Rivers}
\affiliation{Blackett Laboratory, Imperial College London, London
SW7 2AZ, U.K. }\email{r.rivers@imperial.ac.uk}
\author{V.\ P.\ Koshelets}
\affiliation{Institute of Radio Engineering and Electronics,
Russian Academy of Science, Mokhovaya 11, Bldg 7, 125009, Moscow,
Russia.}\email{valery@hitech.cplire.ru}
\date{\today}

\begin{abstract}
It has been argued by Zurek and Kibble that the likelihood of
producing defects in a continuous phase transition depends in a
characteristic way on the quench rate. In this paper we discuss an
improved experiment for measuring the Zurek-Kibble scaling exponent
$\sigma $ for the production of fluxons in annular symmetric
Josephson Tunnel Junctions. We find $\sigma \simeq 0.5$. Further, we
report accurate measurements of the junction gap voltage temperature
dependence which allow for precise monitoring of the fast
temperature variations during the quench.
\end{abstract}

\maketitle

\section{Introduction}

The experiments that we describe here are part of a programme
\cite{Monaco1,Monaco2,Monaco3} designed to see whether continuous
phase transitions in superconductors proceed as fast as they can,
in the sense that the resulting domain structure reflects causal
horizons. Although, under adiabatic change, correlation lengths do
diverge at the critical temperature $T_c$, in reality causality
prevents any lengths diverging since transitions take place in a
finite time. This seemingly simple observation  is not of purely
academic interest, since causality is universal, and the same is
equally true for the early universe with its anticipated rich
sequence of symmetry breaking. Constraints imposed by causal
horizons in the early universe are known to have observable
consequences \cite{kibble0} and, in addition to being of interest
in its own right, one motive for our, and related, work is to see
how transitions in condensed matter systems can mimic this
behaviour.

\noindent The argument for the domain structure after a transition
being determined by causality was first made by Kibble
\cite{kibble1} for the early universe and, independently, by Zurek
\cite{zurek1,zurek2} for condensed matter systems, in what we term
the Kibble-Zurek (KZ) scenario. In particular, if transitions are
frustrated, topological defects (vortices, monopoles, etc.) arise
to mediate the different groundstates of the theory. Their density
will be related to the nature of the domain structure present and
be constrained by causality in turn. The relevance of this is that
defects are, in principle, readily countable, permitting us to
check this proposition. Without taking the parallels any further,
we note that all reasonable models for the early universe show
frustration on cooling \cite{mairi}.

\noindent We would expect that, the faster the quench through the
transition, the more defects we would see.  Whatever the details,
and there are several ways to estimate the defect density in the KZ
picture, all predict characteristic scaling behaviour in the quench
time $\tau_Q$ (the inverse quench rate) defined by:

\begin{equation}
\frac{T_{C}}{\tau _{Q}}=-\frac{dT}{dt}\mid _{T=T_{C}}. \label{tau_q}
\end{equation}

Specifically, if ${\bar\xi}$ is the separation of defects at the
time of their production then it scales with $\tau _{Q}$ as
\begin{equation}
\bar{\xi}\approx\xi _{0}\bigg(\frac{\tau _{Q}}{%
\tau _{0}}\bigg)^{\sigma }.  \label{xibar}
\end{equation}
where $\xi_0$ is, most simply, the cold correlation length, and
$\tau_0$ the relaxation time  of the long wavelength modes.

\noindent The scaling exponent $\sigma >0$ is, in the mean-field
approximation, determined from the static mean-field scaling
exponents and whether the dynamics is largely wavelike (early
universe) or dissipative (condensed matter). That is, universality
classes of continuous adiabatic transitions lead to identical
scaling behaviour of domains as the cooling rate is changed.

\noindent For Josephson Tunnel Junctions (JTJs) the topological
defect is a fluxon i.e. a supercurrent vortex carrying a single
quantum of magnetic flux $\Phi_0 = h/(2e)$ in the plane of the
oxide layer between the two superconductors that make up the JTJ.
Our experiments, summarised in part in \cite{Monaco3}, that we
detail below, show that the spontaneous production of fluxons at a
temperature quench does, indeed, scale with the quench rate,
validating the basic picture of Kibble and Zurek.

\noindent This is a considerable achievement since, in general, it
has proven surprisingly difficult to establish the scaling
behaviour of Eq. (\ref{xibar}). In fact, although there have been
many experiments that have been performed to test the KZ picture
that are commensurate with it or, when not, explicable in its
framework
\cite{chuang,digal,lancaster,lancaster2,grenoble,helsinki,technion,technion2,carmi,Kirtley,
pamplona} our experiments  are the only experiments with condensed
matter systems to date that are sensitive enough to show
unambiguous scaling behaviour.

\noindent It is worth commenting briefly on how using Josephson
Junctions enables us to avoid the main problems that have befallen
the other experiments with superfluids and superconductors which,
superficially, look simpler and more direct than ours.  The first
problem, which besets superfluids, is that of relating the density
of defects at the time of measurement to the density at the time
of formation, as (\ref{xibar}) requires. This was a particular
problem for superfluid $^4He$, leading to the null experiments of
\cite{lancaster2} after the spurious results of \cite{lancaster}.
Although this is avoided in spontaneous vortex production in
superfluid $^3He$, because of the ability to count vortices more
accurately in this case, the fact that $^3He$ is heated by its
disintegration under bombardment by soft neutrons
\cite{grenoble,helsinki} means that the fixed rate of the nuclear
reaction constrains the effective cooling rate. The penalty is
that it is not possible to extract scaling exponents explicitly.

\noindent Both of these problems look to be avoided with planar
superconductors, where flux is conserved, and where a wide range of
cooling rates can be implemented \cite{technion,technion2}. However,
since only net flux can be measured, effectively halving the scaling
exponent, its expected value is so small that it takes an even wider
range of quench rates than are available in order to show scaling
robustly. This is compounded by the fact that, not only are the
systems noisy, but Hindmarsh and Rajantie \cite{rajantie} have shown
that the freezing in of pure magnetic flux gives an additional
contribution to the flux density anticipated from Eq.(\ref{xibar}),
which complicates the issue.

\noindent While JTJs possess the virtues of superconductors, they
avoid those problems since our samples are so small that we expect
to see no more than one (conserved) fluxon in each annulus in each
measurement. That is, for annuli of circumference $C$, we have
${C<\bar\xi}$. As a result, noise is minimised. Also, because JTJs
act as their own thermometers, $\tau _{Q}$ can be measured to high
accuracy. Further, there is no counterpart of the
Hindmarsh-Rajantie mechanism for flux produced in a narrow slit.

\noindent However, we do have a potential problem in that
Eq.(\ref{xibar}), couched in the language of causal horizons, is
not designed for systems, such as ours, that are much smaller than
them. Instead, we propose that the probability $f_{1}$ for
spontaneously producing one fluxon in the thermal quench of a
symmetric annular JTJ of circumference ${C<\bar\xi}$ should scale
with the quench time $\tau _{Q}$ (the inverse quench rate) as

\begin{equation}
f_{1}\simeq \frac{C}{\bar{\xi}}=\frac{C}{\xi _{0}}\bigg(\frac{\tau _{Q}}{%
\tau _{0}}\bigg)^{-\sigma },  \label{P1}
\end{equation}

\noindent where the scaling exponent $\sigma$ depends on the
nature of the junction. We shall provide a justification for
(\ref{P1}) in our concluding sections but, for the moment, we take
it as a consequence of (\ref{xibar}).

\noindent We conclude this section by putting our new experiments
in the context of our old. In 2001 our first proof-of-principle
experiment with JTJs was performed \cite{Monaco1,Monaco2}, to test
the scaling behaviour of (\ref{P1}). The experiment consisted of
taking an annular JTJ isolated from its LHe surroundings and
making it undergo a forced phase transition by heating it above
its superconducting critical temperature and letting it cool
passively back towards the LHe temperature with no external
current or magnetic field. In common with all our experiments, any
trapped fluxon can be detected by the appearance of a current peak
in the I-V characteristic of the JTJ, as will be explained later.

\noindent To derive $\sigma$ two of us had argued earlier
\cite{KMR&MRK} that the relevant causality is provided by the
finite velocity of electromagnetic waves in the JTJ, the Swihart
velocity \cite{Swihart,Barone}. Under the idealistic  assumptions
of a) weak coupling of the superconductors and b) exact critical
slowing of the Swihart velocity at the critical temperature
$T=T_{c}$, we predicted $\sigma = 0.25$\cite{KMR&MRK}. The
experiment was successful, with $\sigma$ commensurate with scaling
behaviour (\ref{P1}) with this value of $\sigma$.

\noindent The experiments performed subsequently, that we shall
describe in this paper, have forced us to revise our assumptions.
In our recent Physical Review Letters \cite{Monaco3} we showed new
scaling behavior for the spontaneous production of fluxons, in
which  $f_{1}$  was indeed seen to clearly follow an allometric
dependence on $\tau _{Q}$, but with a scaling exponent $\sigma =
0.5$. We discuss this experiment in more detail in the subsequent
sections and confirm this behaviour with data from new samples
(see Fig.~\ref{KZData}). The discrepancy between the observations
of the early and late experimental values of $\sigma$ may be less
than it looks at first sight, given the high accuracy of the
latter and the relative scatter of the former. However, from a
theoretical viewpoint, it may also lie in the fact that we need to
take into account the consequences of finite size and the nature
of the fabrication of the junction, which differs between the two
sets of samples. In this regard realistic condensed matter systems
cannot match the early universe for their extension and
uniformity, upon which (\ref{xibar}), with $\sigma$ based simply
on the usual critical exponents, is predicated. Again we postpone
a proper discussion of these issues, and the calculation of
$\sigma$, to the concluding sections of this paper.

\noindent However, before then we shall describe the nature of the
samples and the experimental setup, the use of the JTJ as its own
thermometer to measure quench times, and then the results
confirming the scaling behaviour of (\ref{P1}).

\begin{figure}[ht]
\begin{center}
\epsfysize=7.0cm \epsfbox{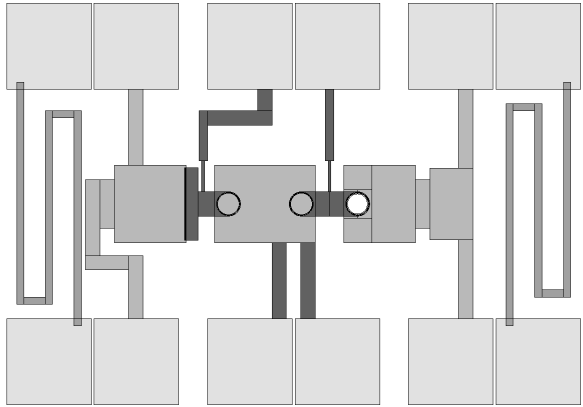}
\end{center}
\caption{Layout of of the $4.2mm \times 3mm$ $Si$ chip containing
four series biased $Nb-Al/Al_{ox}/Nb$ Josephson tunnel junctions. It
integrates three ring shaped junctions having a mean circumference
$C=500\,\mu m$ and a width $\Delta r=4\,\,\mu m$, one $4\times
500\,\mu m^{2}$ overlap-type linear junction and two meanderline
resistive $Mo$ strips used for heating.} \label{layout}
\end{figure}

\section{The samples}

\noindent To begin with some generalities, the annular JTJs
(AJTJs) that we have used are high quality $Nb/Al-Al_{ox}/Nb$ JTJs
fabricated on silicon substrates using the trilayer technique in
which the junction is realized in the window opened in a $SiO$
insulator layer. The so called ''idle region'', i.e. the
overlapping of the wiring layer onto the base electrode was about
$3\,\mu m$ for all the junctions. The thicknesses of the base, top
and wiring layer were $200$, $80$ and $400\,nm$, respectively.
Details of the fabrication process can be found in Ref.\cite{VPK}.
The samples were fabricated at the Superconducting Electronics
Laboratory of the Institute of Radioengineering \& Electronics of
the Russian Academy of Science in Moscow, while the measurements
were carried out at the Physics Department of the Danish Technical
University in Lyngby(DK). All the experiments performed to date
have been carried out on AJTJs with a mean circumference
$C=500\,\mu m$ and a width $\Delta r=4\,\,\mu m$. AJTJs with
larger circumferences have been fabricated, but we have yet to use
them.

The new chip layout developed for the K-Z experiment is shown in
Fig.~\ref{layout}. It integrates three ring shaped junctions having
a mean circumference $C=500\,\mu m$ and a width $\Delta r=4\,\,\mu
m$ and one $4\times 500\,\mu m^{2}$ overlap-type linear junction.
The four JTJs are biased in series. The rightmost AJTJ was obtained
by the superposition of two superconducting rings, as depicted in
Fig.~\ref{geometries}a, while the two AJTJs in the layout middle
were realized by the superposition of a ring shaped top electrode
over a superconducting plane, as shown in Fig.~\ref{geometries}b.
Here we anticipate that this difference in the sample topology did
not affect the measured spontaneous defect production.

For the new experiment a faster and more reliable heating system was
required. This was achieved by integrating a meander line $50\mu$m
wide, $200$nm thick, and $8.3$mm long $Mo$ resistive film in either
ends of the 4.2mm$\times$3mm$\times$0.35mm Si chip containing the
Nb/AlOx/Nb trilayer JTJs (also shown in Fig.~\ref{layout}). These
resistive elements have a nominal d.c. resistance of $50\Omega$ at
LHe temperatures and, due to their good adhesion with the substrate,
are very effective in dissipating heat in the chip. In fact, voltage
pulses a few $\mu$s long and a few volts high applied across the
integrated heater provided quench times as low as $1ms$, that is
more than two orders of magnitude smaller than for the previous
situation \cite{Monaco1,Monaco2}. The on-chip heaters are able to
sustain thousands and thousands of pulses as well $1V$ continuous
bias without any appreciable change of their electrical resistance.

\begin{figure}[htp]
\centering \subfigure[ ]{
    \includegraphics[width=6cm]{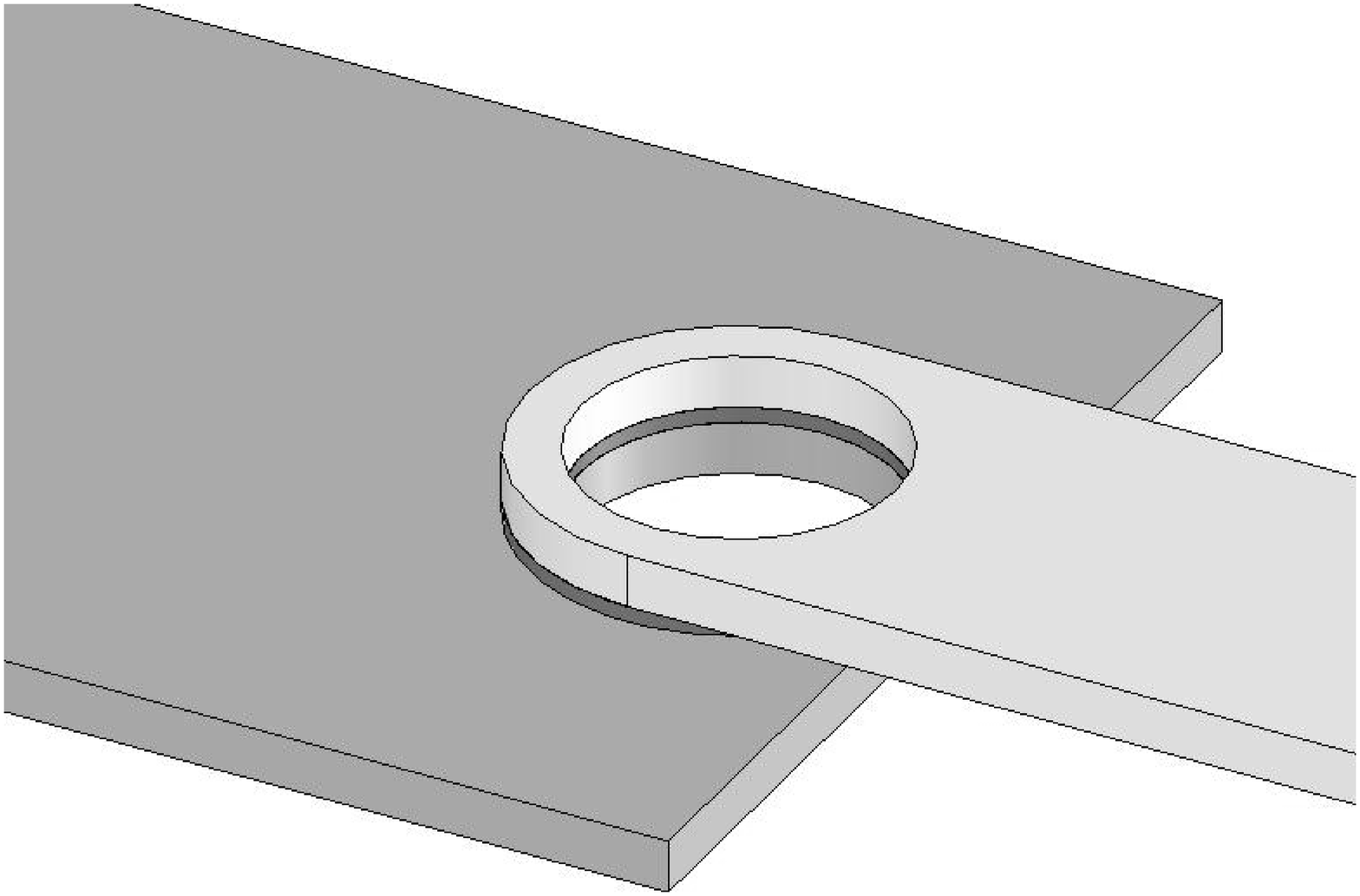}
    \label{geometries}}
\subfigure[ ]{
    \includegraphics[width=6cm]{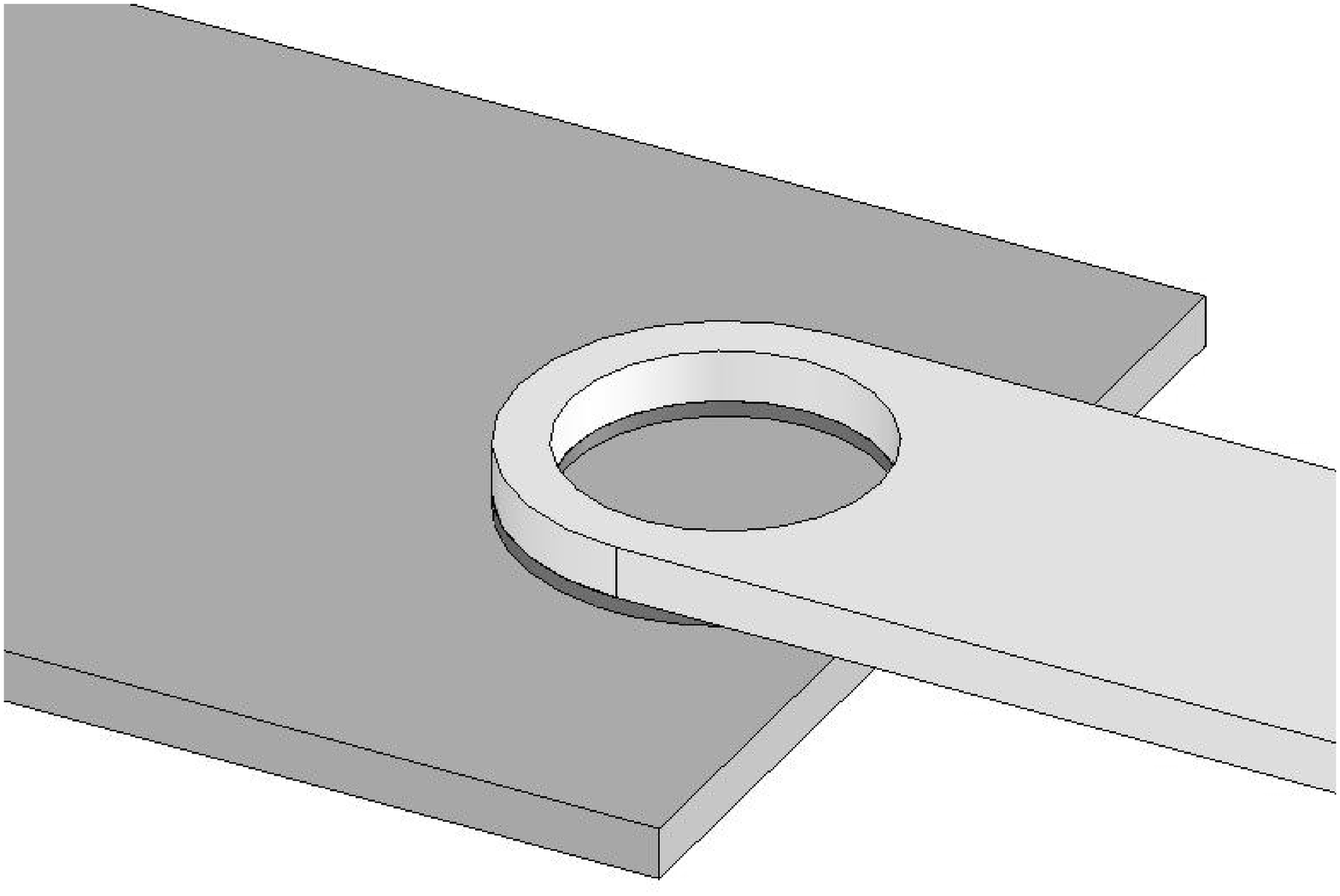}
    \label{geometries}}
\caption{\label{geometries} Sketch not in scale of two geometries
for the annular Josephson tunnel junctions used for the Kibble-Zurek
experiment. The junction base and top electrodes a shown in dark and
light gray, respectively. In (a) the annular junction is obtained by
the superposition of two superconducting rings, while in (b) it is
realized by the superposition of a ring shaped superconducting top
electrode over a superconducting plane base electrode.}
\end{figure}

\section{The experimental setup}

\noindent A detailed description of the experimental setup has been
given in Ref.\cite {Monaco2}. Briefly, the chip with the AJTJs is
mounted to a $Cu$ block enclosed in a vacuum-tight can immersed in
the liquid $He$ bath. A $Ge$ thermometer anchored to the $Cu$ block
allowed for precise measurements of the chip static or slowly
changing temperature. The junction itself is used for quickly
changing temperature measurements, as it will be explained in the
next section, but with lower accuracy. The chip was heated above the
AJTJ critical temperature by a voltage pulse applied to one or both
of the integrated meanderline heaters. Then the heat is removed from
the chip both through the thermal contact with the $Cu$ block and by
$He$ exchange gas inside the can. For all the experiments performed
so far the $He$ exchange gas pressure was fixed at a value of about
$7 mbar$ which makes the heat flow through the copper base plate
predominant.

\noindent In order to minimize thermal gradients during the thermal
cycles, particular care was taken to have clean and flat contact
surfaces of both the chip and the $Cu$ block; further, the voltage
pulses were applied simultaneously to both the on-chip heaters. A
tiny layer of low temperature grease was span underneath the chip to
improve the thermal contact.

\noindent At the end of each cycle the possible spontaneously
generated fluxons are static. An external current supplied to the
AJTJ sets the fluxons (if any) in motion around the annulus and
quantized voltages develop across the junction itself. In other
words, we count the number of produced fluxons by a careful
inspection of the junction current-voltage characteristic(IVC).
(Due to the annihilation of a fluxon-antifluxon pair, this idea
works well as long as the chances to spontaneously generate two
fluxons are small.) Fluxon motion at 4.2K is very unstable in our
samples due to very low junction losses, therefore the IVC was
better inspected at higher temperatures where larger losses
stabilize the fluxon motion. For our sample the optimal
temperature to look for fluxons moving around the ring was in a
range between 6 and 7K.

\noindent Quenching experiments were carried out in a double $\mu
$-metal shielded cryostat and the sampleholder can was surrounded by
a superconducting Pb shield. In turn the chip holder inside the can
was enclosed in a in a cryoperm shield and in one more
superconducting Pb shield. The transitions from the normal to the
superconducting states were performed with no current flowing in the
heaters and the thermometers.  During the quench the JTJ was also
electrically isolated: in fact, both the junction voltage and
current leads were open during the whole thermal cycle. Finally, all
the measurements have been carried out in an electromagnetically
shielded environment.

In order to run batches of several thousand equal thermal cycles
with given parameters, automatization of thermal cycles was
implemented by means of a switching unit controlled by a GPIB
interface; that also allowed for much more robust statistics to be
achieved. At the end of each thermal quench the junction IVC is
automatically digitally acquired, converted and stored. Then at the
end of each batch with a given value of $\tau _{Q}$ an algorithm has
been developed for the analysis of the large amount of IVCs and the
automatic detection and counting of the spontaneously trapped
fluxons.

\section{Determining the quenching time}

The quench time $\tau _{Q}$ was continuously varied over more that
four orders of magnitude (from $20$s down to $1 ms$) by varying
the width and the amplitude of the voltage pulse across the
integrated resistive elements. In order to estimate the quenching
time $\tau _{Q}$ we use the observation by Thouless
\cite{thouless} that the junction itself acts as a thermometer, as
far as its temperature is below the critical temperature. This
gives us an unrivalled accuracy in measuring $\tau_Q$ over other
experiments looking for KZ scaling behaviour. Specifically, the
temperature dependence of the gap energy $\Delta$ in a
strong-coupling superconductor, such as $Nb$,

\begin{equation}
\frac{\Delta (T)}{\Delta (0)}=\tanh \frac{\Delta (T)}{\Delta (0)}\frac{T_{c}%
}{T},\label{gap}
\end{equation}

\noindent also applies to the junction gap voltage $V_{g}$ that is
proportional to $\Delta$. An experimental demonstration of
Eq.(\ref{gap}) in Nb/Nb tunnel junctions having native oxide was
first evidenced in 1976 \cite{broom}, well before the development of
the trilayer technique used here that, by exploiting the more
compact Al oxide, allows for a higher quality and a more stable
tunnel barrier.

Eq.(\ref{gap}) can be manipulated: considering that $arctanh x =
\frac{1}{2} \ln \frac{1+x}{1-x}$ as far as $x^2 <1$ and
approximating $ln (1\pm x) \simeq \pm x - x^2 /2 \pm x^3 /6$, we
easily get:

$$ \frac{\Delta (T)}{\Delta (0)} \simeq \left[ 1-\left(
\frac{T}{T_{c}}\right) ^{4}\right]^{2/3}. $$

\begin{figure}[t]
\begin{center}
\epsfysize=7.0cm \epsfbox{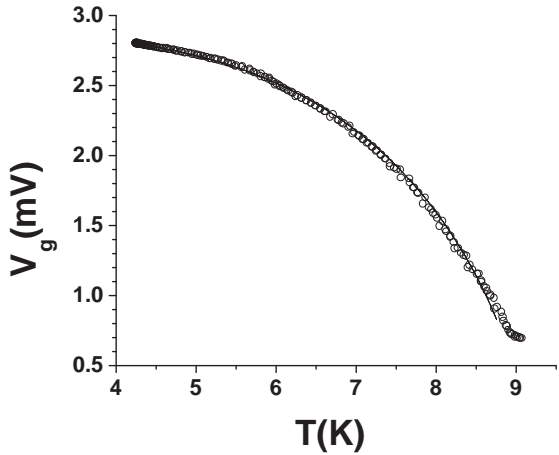}
\end{center}
\caption{The temperature dependence of gap voltage $V_{g}$. The open
circles are the experimental data with the junction biased at 25\%
of the total current jump at the gap voltage; the solid line is the
prediction that follows from Eq.(\ref {gap}). The fit yields
$T_{C}=(9.12\pm 0.04)$K and $V_{g}(T=0)=(2.89\pm 0.02) $mV.}
\label{Vg_vs_T}
\end{figure}

\noindent In our samples we have found that, provided the JTJ is
current biased at about 20-25\% of the total current jump at the gap
voltage, the junction temperature could be monitored to a high
degree of accuracy and speed by resorting to Eq.(\ref{gap}). This is
shown in Fig.~\ref{Vg_vs_T} where the experimental values (open
circles) of the junction gap voltage $V_g(T)=2 \Delta(T) /e$ are
plotted at different values of the temperature as measured by a
calibrated $Ge$ thermometer. The solid line is the prediction that
follows from Eq.(\ref{gap}) with fitting parameters $T_{C}=(9.12\pm
0.04)K$ and $V_{g}(0)=(2.89\pm 0.02)mV$, i.e. $2 \Delta(0)/
k_{B}T_{c} \simeq 3.73 \pm 0.03$. As a result, Eq.(\ref{gap}) can be
used efficiently to estimate the junction temperature to a high
degree of accuracy for $V_{g}(T)>1mV$, i.e., $T<8.5K$. Above $8.5K$
the experimental data for the gap voltage saturate to a finite
temperature-independent value corresponding to the product of the
JTJ normal resistance and the bias current. This way of monitoring
the system temperature is particularly convenient when the
temperature changes rapidly and a complete thermal cycle occurs on a
ms time scale or even shorter. However, the overall temperature
accuracy $\delta T=| \frac{dT}{dV_{g}}|\delta V_{g}$ cannot be
smaller than $2mK$ due to a voltage accuracy $\delta V_{g}$ of about
$2\mu V$ on a fast digital oscilloscope.

Assuming the chip exchanges heat mainly through a massive copper
base plate (with thermal constant $\tau _{1}$) which in turn
exchanges heat with the surrounding helium gas (with thermal
constant $\tau _{2}$), the thermal relaxation during the thermal
quench has been fitted by a double exponential decay of the form:

\begin{equation}
T(t)=T_{fin}+\Delta T_{1} \exp \left( -\frac{t -t_0}{\tau _{1}}%
\right) +\Delta T_{2} \exp \left( -\frac{t -t_0}{\tau _{2}} \right)
, \label{relaxation}
\end{equation}

\noindent with $\Delta T_{1}$, $\Delta T_{2}$, $\tau _{1}$ and $\tau
_{2}$ fitting parameters. $T_{fin}$ and $t_{0}$ are known from the
experiments. If the time origin is triggered by the voltage pulse,
then $t_{0}$ corresponds to the time at which the pulse ends. Once
the parameters in Eq.(\ref {relaxation}) are determined or measured,
the quenching time $\tau_{Q}$ can be inferred from its definition
Eq.(\ref {tau_q}), after Eq.(\ref{relaxation}) has been extrapolated
up to the critical temperature $T_{C}$. At the end of this process
of fitting and extrapolation, $\tau _{Q}$ is known to an overall
accuracy of about $\pm 10\%$.

\section{The K-Z measurements}

\begin{figure}[b]
\begin{center}
\epsfysize=7.0cm \epsfbox{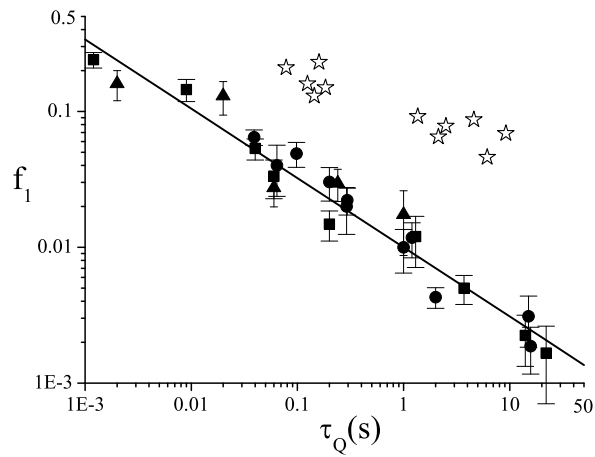}
\end{center}
\caption{Log-log plot of the measured frequency $f_{1}$ of trapping
single fluxons versus the quenching time $\tau _{Q}$. Each point
corresponds to many thermal cycles, closed squares for sample 22-4,
closed triangles for sample 08-3 and closed circles for sample 08-4.
The vertical error bars gives the statistical error while the
relative error bars in $\tau _{Q}$ amounting to $\pm10\%$ are as
large as the dots' width. The solid line is the fit of all data to
an allometric relationship $f_{1}=a\,\tau _{Q}^{-b}$ which yields
$a=0.01\pm 10\%$ (taking $\tau _{Q}$ in seconds) and $b=0.51\pm
5\%$. The open stars represent the data obtained in a previous
experiment\cite{Monaco1,Monaco2}.} \label{KZData}
\end{figure}

The experimental results reported here refer to three identical
AJTJs belonging to two different chips made within the same batch
having a critical current density $J_{c}(0)\simeq 60A/cm^2$ yielding
a Josephson penetration depth $\lambda_{J}(0)\simeq 50\mu m$.
Assuming $\alpha'=\alpha=3.5$ in Eq.(\ref {csi_not}), this leads to
a value of $\xi _{0} \simeq 17 \mu m$. For all samples the high
quality has been inferred by a measure of the I-V characteristic at
$T=4.2\,K$. Due to the very high reliability of the fabrication line
the sample tunnel barriers have the same geometrical and electrical
parameters. However, to distinguish them we will name them after
their chip and junction numbers, that is, we have samples 08-3, 08-4
and 22-4. Samples 08-3 and 08-4 belonged to the same chip. Samples
08-4 and 22-4 had the geometry sketched in Fig.~\ref{geometries}a,
while sample 08-03 had the geometry shown in Fig.~\ref{geometries}b.
The solid symbols in Fig.~\ref{KZData} show on a log-log plot the
measured frequency $f_{1}= n_{1} /N$ of single fluxon trapping,
obtained by quenching the samples $N$ times for each value of a
given quenching time $\tau _{Q}$, $n_{1}$ being the number of times
that the inspection of the low temperature AJTJ current-voltage
characteristics at the very end of each thermal cycle showed that
one defect was spontaneously produced. $N$ ranged between $100$ and
$5000$ and $n_{1}$ was never smaller then $10$, except for some of
the rightmost points ($\tau _{Q}>10s$)for which $n_{1}\geq 3$. All
samples has undergone a total of more than 100,000 thermal cycles
without any measurable change of their electrical parameters. The
vertical error bars gives the statistical error $f_{1}/\surd n_1$.
The relative error bars in $\tau _{Q}$ amounting to $\pm10\%$ are as
large as the dot's width.

\noindent It is quite evident that the dependence of the trapping
frequency on the quenching time is the same for all three AJTJs
independently on the geometry of their base electrodes. Careful
measurements of the junction IVC during the N-S transition indicate
that the junction critical temperature might differ from the
critical temperature of the base electrode film by no more than
$10mK$, while the critical temperature of the wiring film, being
twice thicker, exceeded the critical temperature of the base
electrode by about $100mK$. In other words, at the time the
Josephson effect is installed, the base electrode is only weakly
superconducting and cannot exercise any shielding effect. This
explains why we have not observed any difference between the data of
junction 08-3 and those of junctions 08-4 and 22-4.

\noindent To test Eq.(\ref{P1}), we have fitted the data of all
samples with the same allometric function $f_{1}=a\,\tau
_{Q}^{-b}$, with $a$ and $b$ as free fitting parameters. A linear
regression of $\log f_{1}$ vs. $\log \tau _{Q}$, represented by
the continuous line in Fig.~\ref{KZData}, yields $a=0.01\pm 10\%$
(taking $\tau _{Q}$ in seconds)and $b=0.51\pm 5\%$. Therefore the
present experiment suggests that the scaling exponent is
$\sigma=0.5$, rather than the value $\sigma=0.25$ suggested by our
earlier attempt\cite{Monaco1,Monaco2}.  For comparison, the data
of this experiment are reported in Fig.~\ref{KZData} as open
stars.

The shift in intercept (or, equivalently, prefactor $a$) between
the two sets of data is to be expected. The AJTJs used in the
first experiment, although of the same geometry as samples 08-4
and 22-4 (see Fig.~\ref{geometries}a), had a Josephson current
density $J_c$ about 60 times larger. As we shall reiterate later,
this means a smaller Zurek length $\bar{\xi} \propto 1/ \sqrt
J_c$, with a correspondingly greater likelihood of observing a
fluxon. In other words, we have moved from a situation in which
$C<\bar{\xi}$ to $C\ll\bar{\xi}$. As we shall see, the value
$a=0.01\pm 10\%$ (taking $\tau _{Q}$ in seconds)
 is 6-7 times larger than the predicted value. As a bound we only expect
agreement in the overall normalization $a$ to somewhat better than
an order of magnitude. Empirically, the different condensed matter
experiments have shown that the ratio $a_{\rm observed}/a_{\rm
predicted}$ varies widely from system to system; $O(1)$ for
superfluid $^3He$ \cite{grenoble, helsinki}, very small for
high-$T_c$ superconductors
\cite{technion2}.

\noindent Although the best fit to the stars alone is
$\sigma=0.25$, when seen in conjunction with our new data, this
value is not so compelling, due to the poor statistics and to the
scattering of the stars.   In retrospect, we do not exclude the
possibility of systematic error in the 2001 experiments. The
trapping frequency $f_1$ increases when the thermal cycling occurs
in an externally applied magnetic field. Therefore an insufficient
shielding of the earth's magnetic field might be the cause of the
systematic shift upwards of $f_1$ for increasing $\tau_Q$ in the
first experiment. For the actual experiment we have taken a lot of
precautions against the possibility of having a significant d.c.
residual magnetic field. In fact, although its absolute value
cannot be measured, we have checked that the measured trapping
frequency did not change i) by rotating the sampleholder in the
horizontal plane (which changes the sample orientation with
respect to the direction of the earth's magnetic field)  and ii)
by rotating the chipholder inside the sampleholder, which is kept
fixed (which changes the sample orientation with respect to the
shields). Furthermore, detailed measurements have been carried out
of the dependence of the trapping frequency on the strength of an
external field applied perpendicular to the junction plane for
different samples and for several values of the quenching time.
Such data will be reported in another paper\cite{next} that will
also discuss the interplay between the Abrikosov vortices produced
in the superconducting junction electrodes and the Josephson
vortices produced in the junction barrier during the thermal
quench. As far as the measurements presented in this work are
concerned we made sure that for each value of the quenching time
the corresponding trapping frequency lies at the minimum.

There is another plausible explanation for the different
experimental findings to which we shall return in the next
section. This is that the samples used in the two experiments,
although constructed in the same fabrication line, might have
slightly different proximity interfaces, which results in quite
different temperature dependence of the critical current density
near the transition temperature.

\noindent Whatever the reasons for the differences, the new data of
Fig.~\ref{KZData} resolve another issue. It could be argued that
what we are really seeing is phase ordering from the individual
superconductors, in which case $f_{1}$ would only be a measure of
the flux trapped in just one of the two superconducting annuli
forming the AJTJ, not including those cases in which flux is trapped
simultaneously in both rings. If we had taken the relevant velocity
to be that of phase ordering in the individual superconductors, as
invoked by Zurek \cite{zurek1,zurek2} when considering the
spontaneous flux produced on quenching annuli of simple
superconductors, at the same level of approximation we would have
predicted $\sigma = 0.25$, which is manifestly not the case.
\medskip

There is, however, one unexplained observation. Unlike in our
earlier experiment, in this experiment we never observed the
simultaneous production of two (or more) fluxons (or antifluxons),
despite the fact that we have detected more than $10,000$ single
fluxons for each sample. This is in contradiction with the
expectation that, extrapolating from an infinite junction, the
probability $f_{2}$ of trapping two homopolar fluxons is
$f_{2}=f_{1}^2 /2$, as far as the productions of a defect in
different ring regions are considered independent events.

\section{Theory}

The value of $\sigma = 0.5$ is obviously in disagreement with our
earlier prediction of $\sigma = 0.25$ given in \cite{KMR&MRK}. Two
possible reasons need to be explored. The first is that our
assumptions of idealised JTJs with weak coupling and exact
critical slowing down of the Swihart velocity at the transition
used in \cite{KMR&MRK} are not valid. The second is that the use
of causality seems suspect when dealing with systems like the
annuli discussed here that are smaller than the KZ causal length
$\bar{\xi}$. That is, how should we treat causal horizons that are
larger than the systems at the time that defects are formed, if we
are using them to define defect separation? Is it the case that
Eq.(\ref{P1}) follows from Eq.(\ref{xibar})?

We address the second problem first. In seeing how Eq.(\ref{P1}) can
arise we are helped by the fact that, since the original KZ papers
were published, there has been considerable analytic and numerical
work performed for ideal systems obeying dissipative equations
\cite{karra,moro,rayulti1,laguna,bettencourt,calzetta,antunes,rajantie},
in particular the time-dependent Ginzburg-Landau equation. Although
there has been no attempt to model Josephson tunnel junctions we can
draw several conclusions for JTJs from the phase transitions in
simple systems, such as superfluids and superconductors, that
suggests that small system size is not a problem in principle. This
suggests a rather different picture from that originally posed by
Kibble and Zurek in \cite{kibble1,zurek1,zurek2}, but with the same
outcome in (\ref{xibar}) for large systems.

As originally posed, the separation of defects is directly related
to the correlation length. By definition, the correlation length is
determined from the large distance behaviour of the correlation
function (and thereby the position of the nearest momentum-space
singularity of the power spectrum in the complex plane). On the
other hand, analytic and numerical work shows that defect separation
is more strongly related to the short-distance behaviour of the
correlation function, since it is this that controls the field
zeroes (in this case modulo $2\pi$) with which simple defects are
associated \cite{halperin}. In the first instance after the
transition this is given by the moments of the power spectrum,
rather than its singularities. Of course, if there is only a single
length in the model at the time the transition is implemented, the
two must be essentially identical, although they play very different
roles \cite{antunes}. Provided thermal fluctuations are controlled,
this is plausible for a dissipative system \cite{ray2}. Field
zeroes, which can mature into defects, initially occur on all,
particularly small, scales. Most annihilate very rapidly. What
drives those that have not annihilated to become the cores of
defects on large scales is the unstable growth of long wavelength
modes, which transfers power to long distance fluctuations at the
expense of small. With fluctuations starting from such small
beginnings, from this viewpoint the fact that, ultimately,
$\bar{\xi} > C$, does not hinder defect formation.

Suppose now that we increase $C$ so that $C > \bar{\xi}$ and we
see fluxons every time. $f_1$ is then not a useful measure and,
instead, we measure total flux i.e. the variance $\Delta n$ in the
net number $n$ of fluxons (i.e the number of fluxons minus the
number of antifluxons). Using the spacing of Eq.(\ref{xibar}) a
random walk in phase along the annulus suggests
\begin{equation}
(\Delta n)^2 \approx \frac{C}{\bar{\xi}} = \frac{C}{\xi _{0}}\bigg(\frac{\tau _{Q}}{%
\tau _{0}}\bigg)^{-\sigma },  \label{n}
\end{equation}
Since $\langle n\rangle =0 $, $(\Delta n)^2 = \langle n^2\rangle =
f_1$ when $f_1 < 1$ is sufficiently small that we can neglect the
possibility of seeing more than fluxon.  As a result we rederive
Eq.(\ref{P1}), but without having had to invoke causality in the
same way. More work is being done, particularly on numerical
simulations with periodic boundary conditions \cite{swarup}, but
for the sake of argument we assume that finite size is not the
reason for the disparities in $\sigma$.

It is a second issue as to whether the time scale for defect
formation, the time it takes for long wavelength modes to evolve
fully, is related to the KZ causal time at which defects are
formed, from which (\ref{xibar}) follows. However, very simple
arguments \cite{karra,moro,rayulti1} show that this is the case,
for relatively weak couplings at least, up to logarithmic
corrections. This enables us to use the KZ scenario to calculate
$\sigma$ even though causality is not, directly, the driving
mechanism.

We now turn our attention to the specific properties of the JTJs.
To reiterate, the theory in Ref.\cite{KMR&MRK} was developed for
JTJs whose electrodes are {\it weak coupling} superconductors for
which the temperature dependence of the critical current density
$J_c(T)$ is given by the Ambegaokar-Baratoff equation\cite{AB}:

\begin{equation}
J_{c}(T)=\frac{\pi }{2}\frac{\Delta (T)}{e\rho _{N}}\tanh \frac{\Delta (T)}{%
2k_{B}T},  \label{JcA&B}
\end{equation}

\noindent where $\Delta (T)$ is the superconducting gap energy and
$\rho _{N}$ is JTJ normal resistance per unit area.
Eq.(\ref{JcA&B}) provides a linear decrease of $J_c$ near $T_c$:
\begin{equation}
J_{c}(T)= \alpha J_{c}(0)\bigg(1-\frac{T}{T_{c}}\bigg),
\label{JcTc}
\end{equation}

\noindent  in which the dimensionless quantity $\alpha$ is
approximately equal to $2 \Delta (0)/k_{B}T_{C}=3.5$. However, our
JTJs are based on $Nb$, a {\it strong-coupling} superconductor,
for which Eq.(\ref{JcA&B}) is not necessarily valid. In practice,
high quality and reproducible barriers are achieved by depositing
a thin $Al$ overlay onto the $Nb$ base electrode which will be
only partially oxidized, leaving a $Nb-Al$ bilayer underneath
having a non-BCS temperature dependence of the energy gap and of
the density of states . The proximity effect in $Nb-Al/Al_{ox}/Nb$
JTJs has been extensively studied and it is known to influence the
electrical properties of the junctions, such as the
current-voltage characteristic and the temperature dependence of
the critical current density. Specifically, the proximity effect
in $S-N-I-S$ junctions is responsible for an otherwise subdominant
temperature dependence of the critical current
density\cite{Rowell&Smith} dominating in the vicinity of $T_c$ as:

\begin{equation}
J_{c}(T) \simeq \alpha' J_{c}(0) \left( 1- \frac{T}{T_c}\right)^2,
 \label{JctProx}
\end{equation}

\noindent where $\alpha'$ is a constant depending on the degree of
proximity. The last equation models the tail shaped dependence of
$J_c$ vs. $T$ near $T_c$; it has been theoretically derived and
experimentally confirmed by Golubov {\it et al.}\cite{golubov} in
1995.

\noindent Assuming that the proximity effect is important for our
samples here then, on rephrasing the arguments of
Ref.\cite{KMR&MRK} with Eq.(\ref{JctProx}) replacing
Eq.(\ref{JcTc}), the Josephson penetration depth $\lambda_{J}(T)$,
which plays the role of the system equilibrium coherence length
$\xi(T)$, now diverges linearly near $T_c$ as:

\begin{equation}
\xi(T)=\lambda _{J}(T)\propto J_{c}(T)^{-1/2} =\xi
_{0}\bigg(1-\frac{T}{T_{c}}\bigg)^{-1}, \label{xeq1prox}
\end{equation}

\noindent where
\begin{equation}
\xi_{0}=\sqrt{\frac{\hbar }{2e\mu _{0}d_{s}\alpha' J_{c}(0)}},
\label{csi_not}
\end{equation}

\noindent $d_s$ being the electrode thickness .

At time $t$ close to the transition, the temperature $T(t)$
satisfies
\begin{equation}
\bigg(1-\frac{T(t)}{T_{c}}\bigg)\simeq \frac{t}{\tau_Q},
\label{TcTc}
\end{equation}
where $T(0) = T_c$. The first assumption in the KZ scenario is
that causality establishes a time $\bar t$ at which domains and
defects form, at which time defect separation is $\xi (T({\bar
t}))$.  For JTJs fluxon separation is, from (\ref{xeq1prox}),
\begin{equation}
\bar{\xi} = \xi(T({\bar t}))=\lambda _{J}(T({\bar t}))=\xi
_{0}\frac{\tau_Q}{\bar t}, \label{xeq1prox2}
\end{equation}
instead of the $\xi_0\sqrt{\tau_Q/\bar{t}}$ behaviour of
\cite{KMR&MRK} that we used in \cite{Monaco1,Monaco2}.

 We have argued  that, although it is instability, rather than direct causality, which drives
scaling behaviour, it arises in a way that is quantitatively
indistinguishable from the KZ scenario.

Thus, even where causality is an inappropriate mechanism we can
still adopt its results, that the {\it earliest} possible time $t$
at which defects could possibly appear satisfies

\begin{equation}
{\dot{\xi}}(\bar{t})=-{\bar{c}}(\bar{t}), \label{causality}
\end{equation}

\noindent where ${\bar{t}}$ is the causal time and $\bar{c}$ is
the Swihart velocity.

\noindent As we said, in Ref.\cite{KMR&MRK} we had assumed that
the Swihart velocity vanished at $T_c$ whereas, for realistic
JTJs, it just becomes very small. Swihart\cite{Swihart} has
demonstrated that for a thin-film superconducting strip
transmission line the solution for the velocity varies
continuously  as one passes through the critical temperature into
the normal state. As a result, we assume $\bar{c}(t)=\bar{c}_{nn}$
near the transition temperature where $\bar{c}_{nn}$ is the speed
of light in a microstrip line made of normal metals. In the case
of a microstrip line made by two electrodes having the same
thickness $d_{s}$ and the same skin depth $\delta$, with $d_{s}<<
\delta$, separated by a dielectric layer of thickness $d_{ox}$ and
dielectric constant $\epsilon$, its value,

$$\bar{c}_{nn} = \frac{2}{\delta} \sqrt { \frac{ d_{ox}
d_{s}}{\epsilon \mu}},$$

\noindent depends on the temperature very weakly, but depends on
the frequency $f$ through $\delta=\sqrt{ \rho / \pi \mu f}$,
$\rho$ being the normal metal residual resistivity.

\noindent  The match $\bar{c}(T_c)= \bar{c}_{nn}$ is certainly
realistic and we still have approximate critical slowing down
insofar as $\bar{c}_{nn}$ is much smaller than the zero
temperature Swihart velocity $\bar{c}_0=\sqrt {d_{ox}/{2
\lambda_{L0} \epsilon \mu}}$, i.e., when the zero temperature
London penetration depth $\lambda_{L0} << \delta^2 / d_s$. For
$300nm$ thick Nb electrodes ($\rho= 3.8 \mu \Omega cm$ and
$\lambda_{L0}=90nm$), $\delta \simeq 1mm$ at say $f= 10 kHz$, so
the last inequality is fully satisfied. At the same frequency, for
a value of the specific barrier capacitance $c_s=\epsilon
/d_{ox}=0.02 F/m^2$ typical of low current density
$Nb-Al/Al_{ox}/Nb$ JTJs, we get $\bar{c}_{nn}=7 \cdot 10^3 m/s$
and $\bar{c}_0= 1.4 \cdot 10^7 m/s$. We stress that $\bar{c}_{nn}$
is finite and order of magnitudes smaller than the zero
temperature Swihart velocity $\bar{c}_0$.

\noindent The solution of the causality equation
Eq.(\ref{causality}) with a non-vanishing Swihart velocity yields
a new expression for the so called Zurek or freeze out time
$\bar{t}= \sqrt{ {\xi_{0} \tau_{Q}/{
\bar{c}_{nn}}}}=\sqrt{\tau_{0} \tau_{Q}}$, where $\tau _{0}=\xi
_{0}/ \bar{c}_{nn}$ ($\tau _{0}=O(1ns)$).

\noindent Inserting the value of ${\bar{t}}$ into
Eq.(\ref{xeq1prox}) we obtain the new Zurek length $\bar{\xi}$:

\begin{equation}
{\bar{\xi}}= \xi({\bar{t}})=\sqrt{\xi_{0} \tau_{Q}  \bar{c}_{nn}}=\xi _{0}\bigg(\frac{\tau _{Q}}{\tau _{0}}%
\bigg)^{1/2}. \label{xiZprox}
\end{equation}

\noindent We reach the important conclusion for realistic JTJs
that the probability $f_{1}$ for spontaneously producing one
fluxon in the quench is still predicted to scale with the quench
time $\tau _{Q}$ according to Eq.(\ref{P1}), but the critical
exponent is now $\sigma =\,0.5$, rather than  $\sigma =\,0.25$.

 By varying $\tau_Q$ in the experimentally achieved four
decade range $1 ms < \tau _{Q} < 10 s$, we get  $10 \mu s <
\bar{t} < 1 ms$ that is always much larger than $\tau_0$; it means
that by the time the Josephson phase 'freezes' the Josephson
effect is well established.

\noindent The freezing temperature $\bar{T} =
T(\bar{t})=T_c(1-\bar{t}/\tau_Q)$ comes out to differ by the
critical temperature itself by an amount $T_c-\bar{T}=T_c
\sqrt{\tau_{0}/ \tau_{Q}}$. Further, in the same $\tau_Q$ range
the normalized freezing temperature $\bar{T}/{T_C}$ at which the
defect is formed is $0.99 < \bar{T}/{T_C} <0.9999$. It would be
really hard, if not impossible, to measure the temperature
dependence of $J_c$ and $\bar{c}$ so close to $T_C$. We have then
to resort to theoretical predictions, that is to
Eq.(\ref{JctProx}) for the temperature dependance of the Josephson
current density $J_c()T$. Further, according to Swihart's
calculations and figures \cite{Swihart}, we are allowed to assume
$\bar{c}(t)= c_{nn}$ near the transition temperature.

\section{Conclusions}

\medskip

\noindent Eq.(\ref{P1}) is amenable to further experimental tests
with AJTJs having different critical current densities $J_c(0)$
and/or circumferences $C$. Such experiments should still show the
critical exponent $\sigma=1/2$, however the prefactor should change
accordingly. A test of the importance is provided by working with
asymmetric $Nb-AlN-NbN$ junctions.

 According to the theory presented in this
paper, the expected dependence of the single fluxon trapping
frequency on the critical current is weak, being proportional to its
fourth root, but critical current densities 20-30 times larger
should produce detectable effects. The change in trapping frequency
will be more easily to observe with larger diameter annuli due to
the linear dependence on $C$ in Eq.(\ref{P1}).

\noindent It is worth commenting here on the effect of the
unavoidable thermal gradients in physical system undergoing a
thermal quench; the Zurek-Kibble scaling law Eq.(\ref{P1}) was
derived assuming that thermal gradients are not a problem. More
precisely, according to the general theory of Ref.\cite{kibble2},
the maximum thermal gradient $\nabla T^{*}$ allowed across a given
system  undergoing a thermal quench at the time of the transition
amounts to $(\bar{T}-T_{C})/ \bar{\xi}$. Below this threshold a
saturation is expected in the spontaneous production of defects. In
particular, in our case, where $\tau_Q$ ranges in the interval $10s
- 1ms$, we have $\nabla T^{*} = T_C/(\bar{c}_{nn} \tau_Q)$ ranges
from $6.5 \mu K/mm$ to $65mK/mm$ across our $160\mu m$ diameter ring
corresponding to a critical value for the maximum temperature
difference $\Delta T_{max} = \nabla T^{*}\times C / \pi \approx 1
\mu K - 10 mK$. As stated before, our voltage accuracy in the
measurements of the gap voltage in contiguous JTJs on the same chip
allows us to resolve temperature differences as small as few
millikelvin; the simultaneous measure of the time dependence of the
gap voltages of contiguous junctions in response to the shortest
heat pulse (i.e. corresponding to $\tau_Q = 1ms$) revealed that the
temperature difference, if any, is less than $10 mK$. Further, the
fact that we have not observed any saturation of the defect
production in a wide $\tau _{Q}$ range should indicate that thermal
gradient still do not affect the defect production in our
experiments. Again, in future experiments devised to understand the
role of the thermal gradients, they can be enhanced by progressively
reducing the thermal coupling between the chip and its surroundings
and/or by supplying the voltages pulses only to one of the two
resistive stripes integrated at the chip extremities. On the
opposite, if needed, it is also possible to further reduce the
thermal gradients by using an annular integrated heater superimposed
on and concentric to the annular JTJ.

In summary, we see this experiment as providing unambiguous
corroboration of Kibble-Zurek scaling over a wide range of quenching
time $\tau_Q$ in accord with our predictions for $Nb-Al/Al_{ox}/Nb$
JTJs. As such, it replaces the experiment reported in
\cite{Monaco1,Monaco2} by being more realistic theoretically and
more sophisticated experimentally.

We stress that to date, we are the only group to have performed
experiments on JTJs and this experiment is the \textit{only} one
to have confirmed the Zurek-Kibble scaling exponent for a
condensed matter system.  We note, however, that scaling has been
observed \cite{florence} in non-linear optical systems, which
satisfy equations of the time-dependent Ginzberg-Landau type.

\smallskip

The authors thank P. Dmitriev, A. Sobolev and M. Torgashin for the
sample fabrication and testing and U.L.Olsen for the help at the
initial stage of the experiment. This work is, in part, supported by
the COSLAB programme of the European Science Foundation, the Danish
Research Council, the Hartmann Foundation, the RFBR project
06-02-17206, and the Grant for Leading Scientific School
7812.2006.2.

\end{document}